\documentclass[]{ceurart}
\usepackage{listings}
\usepackage{booktabs}
\usepackage{amsmath}
\usepackage{subcaption}
\newenvironment{rqbox}{%
  \vspace{1pt}\begin{quote}\footnotesize%
}{\end{quote}\vspace{1pt}}

\begin{document}

\copyrightyear{2026}
\copyrightclause{Copyright for this paper by its authors.
  Use permitted under Creative Commons License Attribution 4.0
  International (CC BY 4.0).}

\conference{Joint Proceedings of the STAF 2026 Workshops: AgileMDE,
  GCM, ICMM, LLM4SE, TTC. Rennes, France, June 29--July 3, 2026}

\title{Benchmarking the Titans: A Multi-Dimensional Empirical Evaluation
  of LLM Code Generation Quality in the .NET Ecosystem}

\author[1]{Seyed Mohammad Mahdi Ghalandarian}[%
  email=mohammadmehdi.ghn@aut.ac.ir,
]
\author[1]{Majid Bazargani}[%
  email=Majidbazargani@aut.ac.ir,
]
\author[1]{Masoumeh Taromirad}[%
  email=m.taromirad@aut.ac.ir,
]
\cormark[1]
\address[1]{Amirkabir University of Technology, Iran}
\cortext[1]{Corresponding authors.}

\begin{abstract}
Evaluating Large Language Model (LLM) code generation quality requires
examining not just whether the generated code is correct, but whether it
is maintainable, efficient, and stylistically sound, all of which are
qualities of direct importance to software engineering practitioners.
Existing benchmarks reduce evaluation to a single Pass@k metric, which
obscures critical trade-offs between functional correctness and structural
quality. A further limitation is the near-exclusive focus on Python,
leaving enterprise-relevant ecosystems such as C\# and .NET without
dedicated evaluation. This paper presents an automated, multi-dimensional
evaluation framework for C\# code generation, applying it to four
state-of-the-art LLMs: GPT, Gemini, Claude, and Grok. We conduct a
controlled experiment across 85 algorithmic tasks derived from HumanEval,
generating and evaluating 340 solutions in total, in which each solution
is assessed across three independent dimensions: functional correctness
via automated unit testing, static code quality via Roslyn AST analysis,
and runtime efficiency via adversarial BenchmarkDotNet profiling. Our
central finding reveals a substantial gap between correctness and quality
attributes (Pearson r = 0.075), demonstrating that Pass@k rankings
systematically misrepresent the full LLM performance profile in software
engineering contexts. We further characterize GPT's bimodal failure
behavior.
\end{abstract}


\maketitle

\section{Introduction}

The integration of Large Language Models into software development
workflows has accelerated rapidly~\cite{svyatkovskiy2020intellicode},
with tools such as GitHub Copilot, Cursor, and Claude Code now widely
adopted in both academic and industrial
settings~\cite{ziegler2022productivity}, placing new demands on
evaluation. Industry reports and empirical studies have documented
recurring challenges with LLM-generated code in practice, including
subtle bugs, security vulnerabilities, and maintainability debt that only
becomes apparent after
integration~\cite{perry2023insecure,pearce2022asleep}. The relevant
question is therefore no longer only whether LLM-generated code passes
tests, but whether it meets the quality standards that practitioners are
expected to maintain.

Current evaluation practices, dominated by benchmarks such as
HumanEval~\cite{chen2021codex} and MBPP~\cite{austin2021mbpp}, reduce
performance to Pass@k, the probability that at least one of $k$ generated
samples passes all unit tests. While Pass@k is well-defined, reproducible,
and widely
used~\cite{chen2021codex,austin2021mbpp,li2022alphacode,zheng2023codegeex},
it is blind to code quality properties that directly affect
maintainability, readability, and long-term engineering cost. For example,
a function with cyclomatic complexity of 19 that passes all tests and a
function with complexity of 2 passing the same tests are indistinguishable
under Pass@k, yet they represent meaningfully different artifacts from an
engineering standpoint. This constitutes the first fundamental limitation
of current LLM code generation evaluation: the reduction of a
multi-dimensional quality problem to a single binary correctness signal.

Another limitation is the near-exclusive focus on Python, as evidenced
by the dominant benchmarks in the
field~\cite{chen2021codex,austin2021mbpp,zheng2023codegeex,li2022alphacode}.
The C\#/.NET ecosystem, heavily used in enterprise software development,
financial systems, and game development, lacks a dedicated evaluation
infrastructure that unifies functional correctness testing, runtime
profiling, and static AST analysis into a single automated pipeline for
LLM-generated code~\cite{paul2024benchmarks,zheng2023codegeex}. While
tools such as NDepend, SonarQube, and Roslynator provide static analysis
capabilities for C\# source code, none integrates all three evaluation
dimensions---correctness, structural quality, and runtime
efficiency---into a unified automated benchmarking harness targeted at
LLM outputs. Practitioners working in .NET environments are therefore
unable to draw reliable, ecosystem-specific conclusions from existing
benchmark results.

This paper addresses both limitations by introducing a fully automated
end-to-end C\# evaluation framework applied to GPT, Gemini, Claude, and
Grok. The framework operates in four sequential phases (dataset
preparation, code generation, dynamic evaluation, and result collection),
integrating functional correctness testing via .NET Reflection,
Roslyn-based\footnote{Microsoft Roslyn is the open-source .NET compiler
  platform that exposes a full API for code analysis and transformation.
  Available at: \url{https://github.com/dotnet/roslyn}} AST static
analysis, and adversarial BenchmarkDotNet~\cite{akinshin2019dotnet}
runtime profiling into a unified harness, with all results consolidated
into a structured dataset for cross-dimensional analysis. Our central
contribution is demonstrating, through this framework, that functional
correctness and structural code quality are empirically orthogonal---a
finding that directly challenges the sufficiency of Pass@k as the sole
evaluation criterion for LLMs in software engineering contexts.

\section{Related Work}
\label{sec:related}

Code generation evaluation has evolved considerably since the introduction
of early function-level benchmarks, expanding from simple correctness
metrics toward richer assessments of repository-level complexity,
evaluation integrity, and code quality. This section situates our work
within these three threads of the literature.

HumanEval~\cite{chen2021codex} and MBPP~\cite{austin2021mbpp} established
Pass@k as the dominant metric for LLM code generation, providing a
reproducible and well-defined correctness signal. However, more recent
work argues that isolated function generation does not reflect real-world
complexity. DevEval~\cite{lian2024deveval} introduced 2,690 samples drawn
from 119 real-world projects, while HumanEvo~\cite{zheng2025humanevo}
demonstrated that existing evaluation methods substantially overestimate
LLM performance at repository level, with performance overestimations
ranging from 10\% to 61\% when evolution-aware evaluation is applied.
Jiao et al.~\cite{jiao2023evaluation} demonstrated that existing
benchmarks are insufficient to evaluate LLMs across different code
translation complexity levels, and Paul et al.~\cite{paul2024benchmarks}
provided a critical review confirming that current metrics fail to capture
the full capability spectrum of LLMs in code generation. Despite these
advances, C\# and the .NET ecosystem remain substantially
under-represented across all existing benchmark suites.

A parallel line of work has focused on evaluation integrity.
Dou et al.~\cite{dou2025wrong} showed that LLM-generated code frequently
contains structural and logical deficiencies beyond simple correctness
failures, with models tending to produce shorter yet more complex code
compared to canonical solutions---a pattern that already suggests
correctness and structural quality need not co-vary. Zhang
et al.~\cite{zhang2024hallucinations} demonstrated that hallucinations
and contextual dependency failures are a primary source of runtime errors
in LLM-generated code, a finding that directly motivates our decision to
build a full .NET solution structure that resolves project-level
dependencies during dynamic compilation.

Closest to our quality-measurement objectives,
TASTY~\cite{moudgalya2023tasty} predicts algorithmic complexity of C++ and
Python code using a Transformer classifier. However, it predicts
complexity classes rather than empirically measuring them, and does not
assess runtime efficiency or naming quality. To our knowledge, no prior
work integrates runtime benchmarking, static analysis, and functional
correctness testing into a single automated pipeline for the .NET
ecosystem.

\section{Methodology and Framework Architecture}
\label{sec:methodology}

This section describes the design and implementation of the evaluation
framework in full detail. The framework assesses LLM-generated C\# code
across three independent quality dimensions: (1)~\textit{functional
correctness}, measuring whether generated solutions pass all unit tests;
(2)~\textit{static code quality}, capturing structural and stylistic
properties via AST analysis; and (3)~\textit{runtime efficiency},
measuring execution time and memory allocation under adversarial
profiling. The research questions below are organized around these three
dimensions.

\subsection{Research Questions}

To structure our empirical investigation, we define the following
research questions:

\begin{description}
  \item[RQ1 (Correctness):] How does functional correctness compare
    across GPT, Gemini, Claude, and Grok when generating C\# solutions?
  \item[RQ2 (Task Complexity):] To what extent does inherent task
    difficulty, rather than model-specific behavior, drive the structural
    complexity (cyclomatic complexity and nesting depth) of generated code?
  \item[RQ3 (Orthogonality):] Is there a statistically significant
    relationship between functional correctness and composite code quality,
    or are these independent evaluation dimensions?
  \item[RQ4 (Performance):] How do LLM-generated C\# solutions compare
    in runtime execution efficiency and memory allocation under adversarial
    BenchmarkDotNet profiling?
\end{description}

\subsection{Evaluation Metrics}

The metrics were selected to directly address each research question.
For RQ1, functional correctness is the Success Rate (\%), i.e., the
proportion of unit tests passed per solution. For RQ2, cross-model
pairwise Pearson correlations on cyclomatic complexity and nesting depth
across identical tasks quantify how much task difficulty---rather than
model behavior---drives structural
complexity~\cite{mccabe1976complexity,mccabe1989design}. For RQ4,
BenchmarkDotNet~\cite{akinshin2019dotnet} yields median Execution Time
(ns) and heap Memory Allocation (bytes/operation). For RQ3, a Composite
Quality Score aggregates five normalized sub-components (correctness,
complexity, nesting, style, and performance) into a single 0--100 scale
value (formula defined in Phase~4).

\subsection{The Evaluation Framework}

\begin{figure}[h!]
  \centering
  \includegraphics[width=0.52\linewidth]{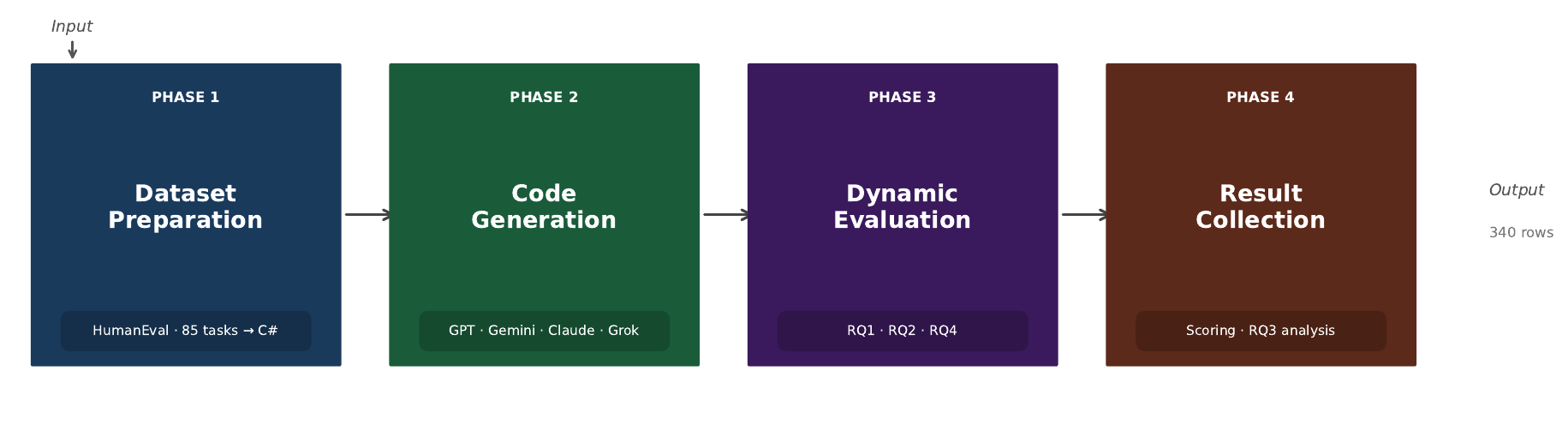}
  \caption{\footnotesize High-level overview of the four-phase evaluation
    pipeline and their mapping to RQ1--RQ4.}
  \label{fig:highlevel}
\end{figure}

The framework was built as a fully automated, end-to-end pipeline that
takes a benchmark dataset as input and produces a structured
multi-dimensional evaluation result for given LLMs under study. The
pipeline consists of four major phases, namely dataset preparation, code
generation, dynamic evaluation, and result collection. The overall
architecture is illustrated in Figure~\ref{fig:highlevel}.

\paragraph{Phase 1: Data Preparation.}
Each HumanEval~\cite{chen2021codex} record contains five fields ---
\texttt{task\_id}, \texttt{description}, \texttt{prompt},
\texttt{entry\_point}, and \texttt{test} --- transformed into a
structured Excel spreadsheet (one authoritative row per problem).
A representative record is shown in Table~\ref{tab:record}; the
\texttt{test} column shows a single representative assertion for
brevity---each task contains multiple assertions from the original
HumanEval test suite.

\begin{table}[ht]
  \centering
  \captionsetup{skip=4pt}
  \caption{\scriptsize Representative prepared record for task
    \texttt{csharp\_1}.}
  \label{tab:record}
  \setlength{\aboverulesep}{0.5pt}
  \setlength{\belowrulesep}{0.5pt}
  \renewcommand{\arraystretch}{0.78}
  \scriptsize
  \begin{tabular}{ll}
    \toprule
    \textbf{Field} & \textbf{Content} \\
    \midrule
    \texttt{task\_id}          & \texttt{csharp/1} \\
    \texttt{description}       & Given two strings s and t, return true if t is an anagram of s \\
    \texttt{prompt}            & \texttt{public static bool IsAnagram(string s, string t)} \\
    \texttt{entry\_point}      & \texttt{IsAnagram} \\
    \texttt{test}              & \texttt{Assert.AreEqual(true, Solver.IsAnagram("anagram","nagaram"))} \\
    \texttt{solution\_prompt}  & Role: Expert C\# Developer, class Solver, match signature, no namespace \\
    \texttt{benchmark\_prompt} & BenchmarkRunner, worst-case anagram inputs, N = 100/1,000/10,000 \\
    \bottomrule
  \end{tabular}
  \vspace{2pt}
\end{table}

\begin{figure}[h!]
  \centering
  \includegraphics[width=0.62\linewidth]{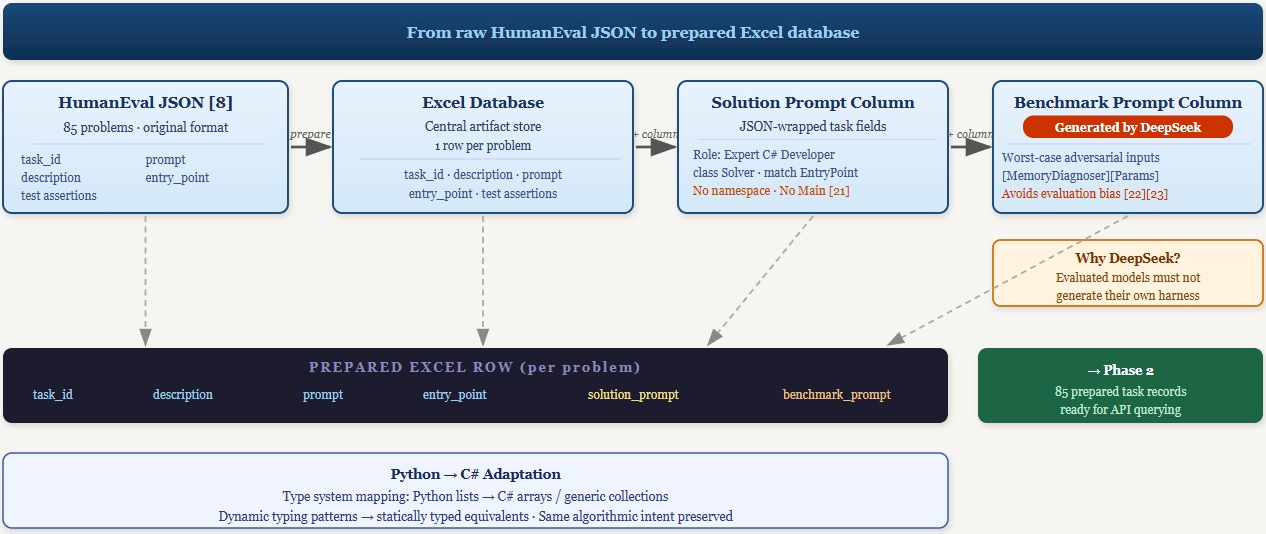}
  \caption{\footnotesize Phase~1: Data Preparation -- HumanEval records
    transformed into an Excel database.}
  \label{fig:dataset}
\end{figure}

Two additional columns were computed per problem. The
\texttt{solution\_prompt} column is the field submitted to each LLM API
on every individual run; it embeds the task fields into an engineered
system prompt that enforces a \texttt{public class Solver} output
structure, prohibiting namespace declarations and \texttt{Main} methods
injected later by the engine---any structural deviation would break
automated compilation~\cite{huang2023empirical}. The
\texttt{benchmark\_prompt} generates a BenchmarkDotNet harness using
worst-case inputs designed to prevent early-exit
optimizations~\cite{kalibera2013rigorous,akinshin2019dotnet}.
Deliberately, harnesses were generated by an independent
DeepSeek~\cite{deepseek2024v3} instance rather than any evaluated model,
preventing input bias and ensuring all four solvers were profiled under
identical conditions~\cite{liang2022helm,chang2024survey}. The prepared
dataset is illustrated in Figure~\ref{fig:dataset}.

\begin{figure}[h!]
  \centering
  \includegraphics[width=0.62\linewidth]{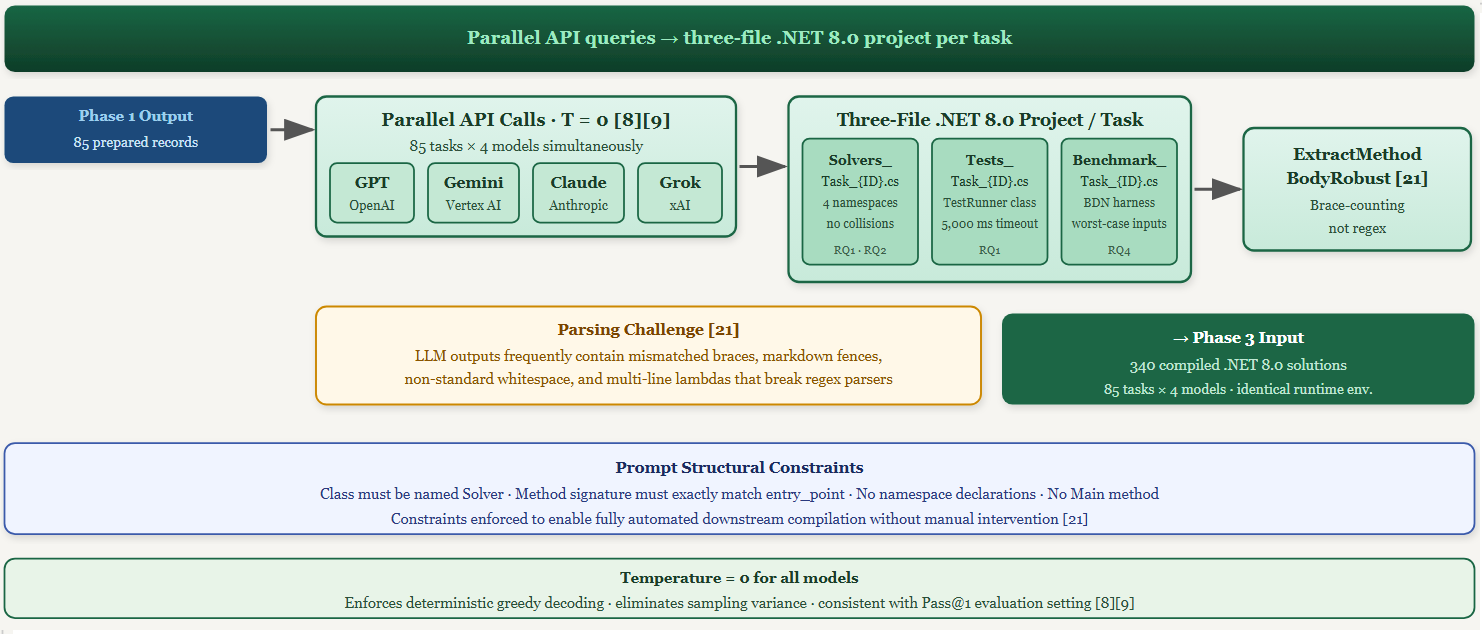}
  \caption{\footnotesize Phase~2: Code Generation -- parallel API queries
    assembled into a shared three-file .NET~8.0 project per task.}
  \label{fig:generation}
\end{figure}

\paragraph{Phase 2: Code Generation.}
The framework queried all four LLMs in parallel via their respective
APIs, submitting the \texttt{solution\_prompt} for each of the 85 tasks
at temperature = 0 to enforce deterministic decoding and eliminate
sampling variance, consistent with the Pass@1 evaluation
setting~\cite{chen2021codex,austin2021mbpp}. Models are detailed in
Section~\ref{sec:setup}. All four outputs per task are compiled into a
shared .NET~8.0 project comprising three files: the generated solutions
(each \texttt{Solver} class isolated in its own namespace), the benchmark
harness, and a test runner with a 5,000\,ms timeout, as illustrated in
Figure~\ref{fig:generation}. Since LLM outputs frequently contain
formatting inconsistencies that cause regex-based parsers to extract
malformed method
bodies~\cite{huang2023empirical,zhang2024hallucinations}, the framework
implements the \textsc{ExtractMethodBodyRobust} algorithm, which uses
brace-counting to correctly locate method boundaries regardless of
nesting depth or formatting variance.

\paragraph{Phase 3: Evaluation.}
The evaluation is carried out in three passes, namely functional
correctness testing, static code quality analysis, and runtime
performance profiling, each addressing a dedicated research question as
defined in Section~3.1.

\textit{1.~Correctness.} Each solver was invoked via .NET Reflection
(model-specific \texttt{Run\_\{AI\}} method), executing the full set of
unit assertions from the HumanEval test field~\cite{chen2021codex}. A
thread-local score accumulator computes the Success Rate (\%) as the
proportion of assertions passed. Deep equality via CompareNetObjects
handles complex return types (lists, arrays) to avoid false negatives.

\textit{2.~Static Quality.} The second pass assessed static code quality
(addressing RQ2) using Microsoft Roslyn, which exposes each solution's
full Abstract Syntax Tree for programmatic inspection. Four metrics are
extracted: \textit{Cyclomatic Complexity}~\cite{mccabe1976complexity},
counting linearly independent execution paths (all \texttt{if},
\texttt{while}, \texttt{for}, \texttt{switch}, \texttt{\&\&},
\texttt{||} constructs), where higher values indicate harder-to-maintain
code; \textit{Nesting Depth}, the maximum depth of nested control
structures, correlated with cognitive
load~\cite{mccabe1976complexity,mccabe1989design}; \textit{Naming
Violations}, counting deviations from standard C\# casing conventions
(PascalCase for public members, camelCase for
locals),\footnote{Microsoft C\# Coding Conventions: \url{https://learn.microsoft.com/en-us/dotnet/csharp/fundamentals/coding-style/coding-conventions}}
which were not stated in the prompt---making this metric a measure of
implicit convention adherence rather than instruction-following; and a
\textit{Professionalism Score} capturing XML documentation comments and
explicit null-checks expected in production C\# codebases.

\textit{3.~Runtime Performance.} The third evaluation pass (addressing
RQ4) subjected all compiled solutions to runtime profiling using
BenchmarkDotNet, one of the most widely used .NET microbenchmarking
libraries. All benchmark methods ($85 \text{ tasks} \times 4
\text{ models}$) were executed in a single joined run. For each task,
the \texttt{BenchmarkRunner} class constructed during the preparation
phase was executed across three input sizes (N = 100, 1,000, and
10,000) using the worst-case data generated by DeepSeek, ensuring that
all execution paths within each solver were fully exercised. Two
performance metrics were recorded per model per task: \textit{median
Execution Time} in nanoseconds and \textit{total heap Memory Allocation}
in bytes per operation. The large iteration count is necessary to
produce stable, low-variance estimates and to amortize the overhead of
.NET's JIT compilation across
measurements~\cite{kalibera2013rigorous,georges2007statistically}.

\paragraph{Phase 4: Result Collection and Composite Scoring.}
Upon completion of all three evaluation passes, the results for all 340
solution--model pairs ($85 \text{ tasks} \times 4 \text{ models}$) were
consolidated into a structured Excel file, with one row per pair and the
following columns: Task ID, AI Model, Correctness (\%), Quality Score,
Time (ns), Memory (Bytes), Complexity, Nesting, and Naming Violations.

To enable the cross-dimensional analysis required by RQ3, a Composite
Quality Score was also computed for each row, aggregating five normalized
sub-components into a single value on a 0--100 scale:

\begin{equation}
  \text{Composite Quality Score} =
    \frac{\text{Norm\_C} + \text{Norm\_Cx} + \text{Norm\_N}
          + \text{Norm\_S} + \text{Norm\_P}}{5} \times 100
  \label{eq:score}
\end{equation}

\noindent where \textbf{Norm\_C} = normalized correctness,
\textbf{Norm\_Cx} = normalized cyclomatic complexity (inverted),
\textbf{Norm\_N} = normalized nesting depth (inverted),
\textbf{Norm\_S} = normalized naming/style score, and \textbf{Norm\_P}
= normalized professionalism score.

Complexity and Nesting are inversely normalized so that simpler, flatter
code scores higher, consistent with established software quality
principles~\cite{mccabe1976complexity,mccabe1989design,lian2024deveval,jiao2023evaluation,dou2025wrong}.
Performance is normalized against a 10,000\,ns baseline using a decay
curve that penalizes disproportionately slow
solutions~\cite{jiao2023evaluation,paul2024benchmarks}. All five
components are weighted equally~\cite{lian2024deveval,dou2025wrong}; we
acknowledge that the equal weighting and decay curve parameters are
design choices that may not generalize, and calibration via expert study
is identified as future work.

\section{Experimental Results}
\label{sec:results}

This section presents the experimental results of the evaluation of four
well-known LLMs, namely GPT-4~\cite{openai2023gpt4},
Gemini~1.5~Pro~\cite{google2024gemini},
Claude~3.5~Sonnet~\cite{anthropic2024claude}, and
Grok~3~\cite{xai2025grok} (OpenAI, Google, Anthropic, and xAI
respectively), applied to the C\# code generation framework described in
Section~\ref{sec:methodology}.

\subsection{Experimental Setup}
\label{sec:setup}

All experiments were conducted on a Windows~11 machine equipped with an
AMD Ryzen~9~5900X processor and 16\,GB RAM, running the .NET~8.0~SDK.
The four models evaluated were GPT-4 (OpenAI API), Gemini~1.5~Pro
(Google Vertex AI), Claude~3.5~Sonnet (Anthropic API), and Grok~3 (xAI
API). All models were queried with API temperature set to 0 to enforce
deterministic decoding, consistent with the Pass@1 evaluation setting
described in Section~\ref{sec:methodology}; all four providers document
greedy decoding at temperature = 0 as deterministic for a fixed model
version, which mitigates, though does not entirely eliminate,
run-to-run variance~\cite{chen2021codex,austin2021mbpp}.
BenchmarkDotNet harnesses were generated independently using
DeepSeek-V3~\cite{deepseek2024v3}, deliberately excluded from the
evaluation pool to prevent bias in worst-case profiling
inputs~\cite{liang2022helm,chang2024survey}. Runtime profiling was
performed using BenchmarkDotNet across three input sizes (\(N = 100\),
\(1,000\), and \(10,000\)) per task per model to ensure stable,
low-variance performance
estimates~\cite{akinshin2019dotnet,kalibera2013rigorous,georges2007statistically}.

\subsection{Results Overview}

Table~\ref{tab:summary} summarises per-model outcomes across all
metrics. RQ1: Sec.~\ref{sec:rq1}; RQ2: Sec.~\ref{sec:rq2};
RQ4: Sec.~\ref{sec:rq4}. RQ3 is addressed jointly: ranking reversal
evidence in Sec.~\ref{sec:orthogonality}; full Pearson analysis in
Sec.~\ref{sec:correlation}.

\begin{table}[ht]
  \centering
  \captionsetup{skip=4pt}
  \caption{\scriptsize Per-model summary (Med.\ time and memory used,
    see Sec.~\ref{sec:rq4}).}
  \label{tab:summary}
  \setlength{\aboverulesep}{1pt}
  \setlength{\belowrulesep}{1pt}
  \setlength{\cmidrulesep}{1pt}
  \renewcommand{\arraystretch}{0.8}
  \scriptsize
  \begin{tabular}{lccccccc}
    \toprule
    \textbf{Model} & \textbf{Correct.} & \textbf{Quality} & \textbf{Cmplx.} &
    \textbf{Nest.} & \textbf{Time (ns)} & \textbf{Mem. (MB)} & \textbf{Fails} \\
    \midrule
    GPT    & 96.37 & 38.31 & 5.16 & 1.68 & 4,823  & 11.7  & 6 \\
    Gemini & 98.53 & 40.05 & 4.12 & 1.62 & 9,034  & 28.9  & 2 \\
    Claude & 98.32 & 37.69 & 2.47 & 1.13 & 12,859 & 156.6 & 4 \\
    Grok   & 97.98 & 40.80 & 3.53 & 1.41 & 9,136  & 6.0   & 4 \\
    \bottomrule
  \end{tabular}
  \vspace{2pt}
\end{table}

\subsection{Functional Correctness (RQ1)}
\label{sec:rq1}

\begin{figure}[ht]
  \centering
  \subcaptionbox{\footnotesize Mean Correctness (\%) with $\pm$1\,SD
    error bars and failure counts per model.\label{fig:correctness}}
    [0.48\linewidth]{\includegraphics[width=\linewidth]{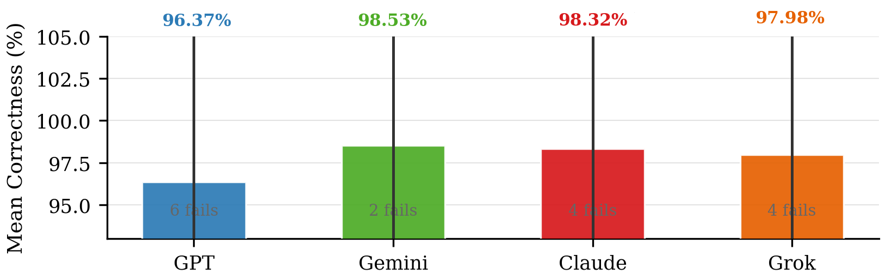}}
  \hfill
  \subcaptionbox{\footnotesize Cyclomatic Complexity and Nesting Depth
    per model. Claude generates the simplest code (2.47 vs
    GPT's 5.16).\label{fig:complexity}}
    [0.48\linewidth]{\includegraphics[width=\linewidth]{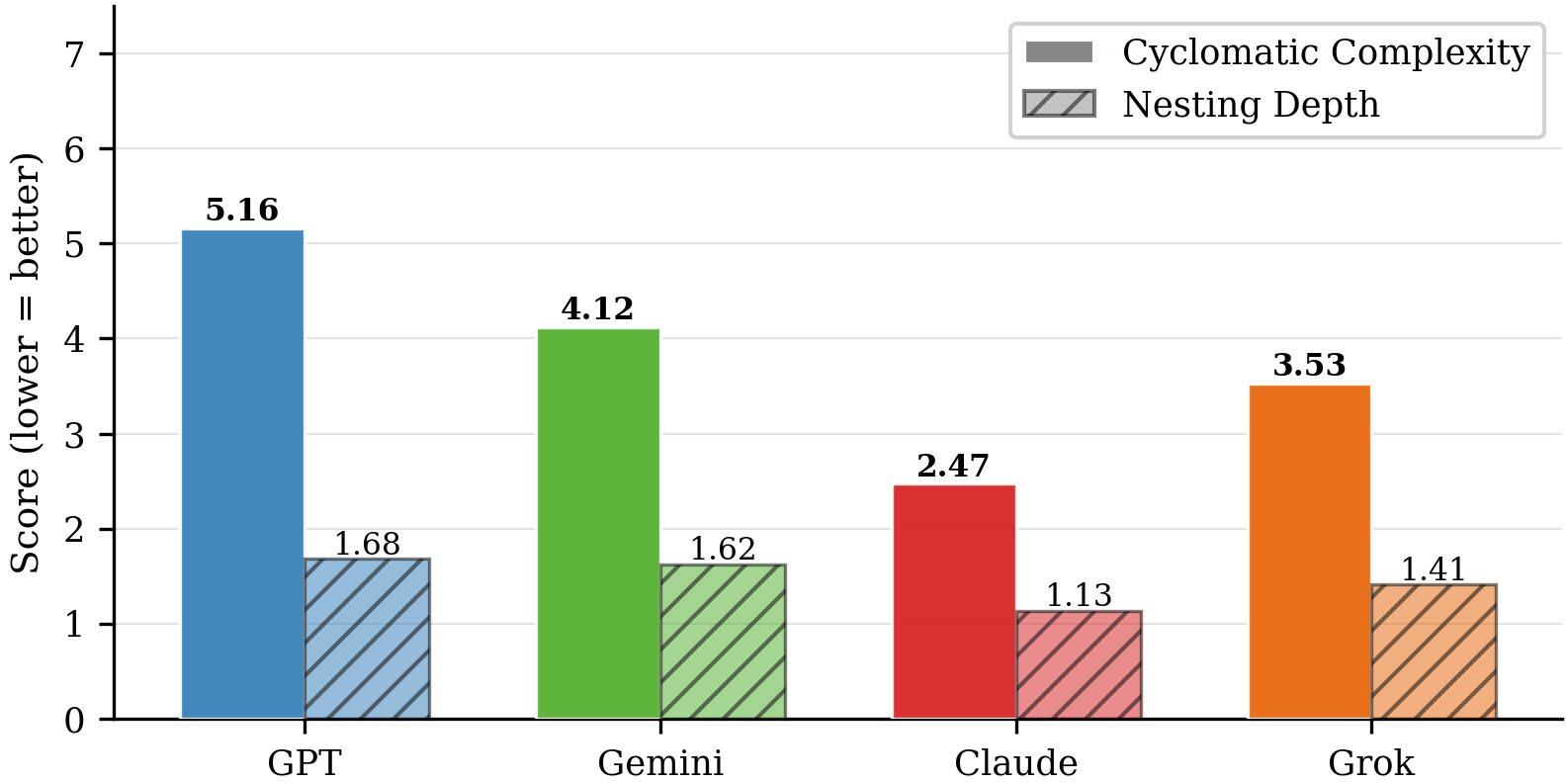}}
  \caption{\footnotesize Evaluation of model performance: Correctness
    rates and code complexity metrics.}
  \label{fig:performance_metrics}
\end{figure}

Addressing RQ1, Figure~\ref{fig:correctness} shows mean correctness with
standard deviation error bars and failure counts. All four models achieve
a median correctness of 100\%, but meaningful differences emerge in mean
and variance. Gemini leads at 98.53\% (2 failures), followed by Claude
at 98.32\% (4 failures), Grok at 97.98\% (4 failures), and GPT at
96.37\% (6 failures) with the highest variance (SD = 15.37 vs Claude's
9.46), indicating a bimodal tendency toward fully correct or substantially
failed solutions. Two tasks were universally challenging:
\texttt{csharp\_57} exposed a shared model limitation (Gemini scored
0\%), and \texttt{csharp\_64} produced an identical 75.0\% across all
four models, suggesting shared training data overlap.

\subsection{Task Complexity and Model Behavior (RQ2)}
\label{sec:rq2}

Addressing RQ2, Figure~\ref{fig:complexity} shows cyclomatic complexity
and nesting depth across models. Claude generates the structurally
simplest code (complexity 2.47, nesting 1.13): its cyclomatic complexity
is approximately half that of GPT (5.16), and its nesting depth is
roughly one third lower (1.13 vs.\ 1.68). However, the more
consequential finding concerns the source of these differences.
Cross-model complexity agreement on identical tasks yields Pearson
correlations of \mbox{\(r = 0.63\) to \(0.69\)} across all model pairs,
indicating that a substantial portion of the complexity observed in any
model's output is attributable to the inherent difficulty of the task
rather than to model-specific generation behavior. Problems requiring
recursive logic or multi-branch control force elevated complexity across
all four models simultaneously.

\subsection{Runtime Performance (RQ4)}
\label{sec:rq4}

\begin{figure}[ht]
  \centering
  \includegraphics[width=0.50\linewidth]{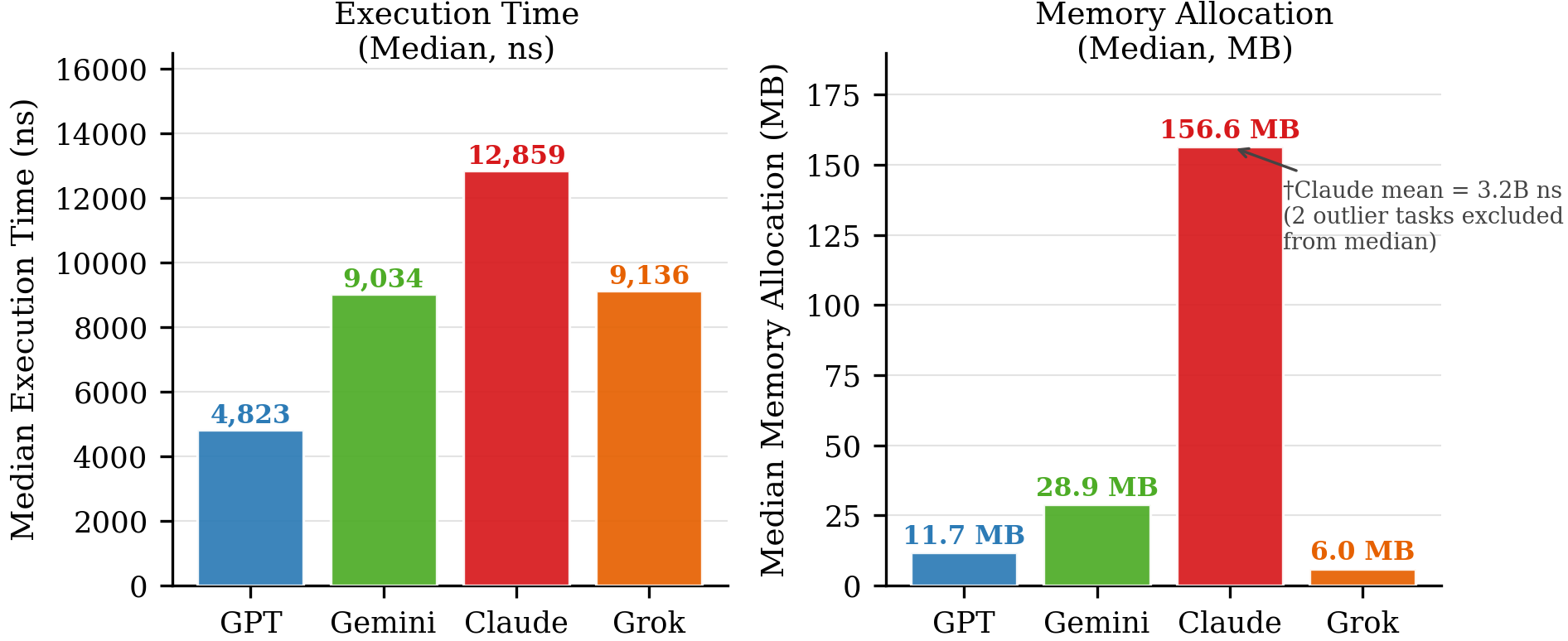}
  \caption{\footnotesize Median execution time (left) and memory
    allocation (right). Median used due to outliers in Claude's data.}
  \label{fig:performance}
\end{figure}

Addressing RQ4, Figure~\ref{fig:performance} shows median execution time
and median memory allocation. Median values are used rather than means
due to extreme outliers in Claude's data. \textbf{GPT} is the fastest at
median (4,823\,ns). Memory allocation varies substantially across models:
\textbf{Grok} allocates the least (6.0\,MB), followed by \textbf{GPT}
(11.7\,MB), \textbf{Gemini} (28.9\,MB), and \textbf{Claude}
(156.6\,MB)---the latter being roughly 26$\times$ higher than Grok and
13$\times$ higher than GPT. \textbf{Claude}'s median execution time
(12,859\,ns) and memory allocation are both the highest, driven primarily
by two anomalous tasks identified during profiling: \texttt{csharp\_12},
in which Claude's execution time reached approximately $2.56 \times
10^{11}$\,ns---roughly 2,500$\times$ slower than the next worst model on
the same task---and \texttt{csharp\_16}, in which Claude allocated
approximately 3.94\,GB of heap memory while all other models allocated
near zero, suggesting a fundamentally different and pathological
algorithmic strategy for these specific tasks.

\section{Discussion}
\label{sec:discussion}

\subsection{Inter-Metric Correlation Analysis}
\label{sec:correlation}

\begin{figure}[h]
  \centering
  \includegraphics[width=0.41\linewidth]{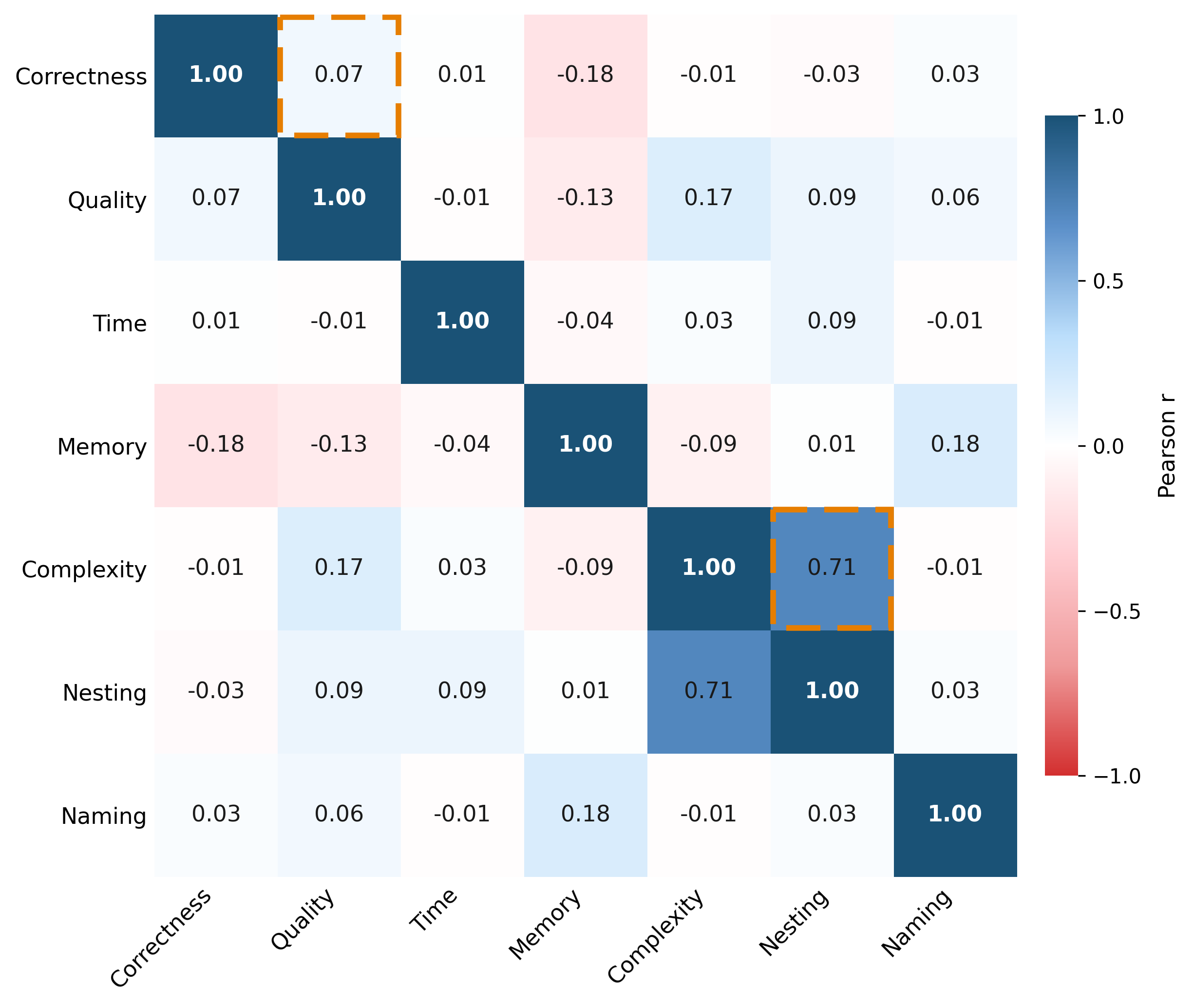}
  \caption{\footnotesize Pearson Correlation Matrix ($n\,=\,340$):
    Complexity--Nesting $r = 0.706$; Correctness--Quality $r = 0.075$.}
  \label{fig:correlation}
\end{figure}

Figure~\ref{fig:correlation} presents the full Pearson correlation
matrix~\cite{kitchenham2002preliminary} across all 340 records. Two
off-diagonal cells are highlighted as the key findings of this study:
the Complexity--Nesting cell ($r = 0.706$) and the
Correctness--Quality cell ($r = 0.075$).

The strongest pairwise correlation is Complexity--Nesting ($r = 0.706$),
confirming that these structural metrics are not independent, as models
generating complex logic also produce deeply nested code. More
consequentially for benchmark design, Correctness is near-zero correlated
with all other metrics ($|r| < 0.18$ throughout), including Quality Score
($r = 0.075$). We note that because Norm\_C (normalized correctness) is
itself one of the five components of the Composite Quality Score
(Equation~\ref{eq:score}), the reported $r = 0.075$ specifically reflects
the near-zero association between raw correctness and the four remaining
structural and performance components, which together pull against any
upward bias. This orthogonality is the central empirical result:
correctness and structural quality must be evaluated as separate,
independent dimensions.

\subsection{The Correctness--Quality Orthogonality Problem}
\label{sec:orthogonality}

The near-zero correlation ($r = 0.075$) means Pass@k rankings are
statistically uninformative about code quality. Prior work has implicitly
assumed that a model capable of generating functionally correct code
would also tend to produce cleaner, more maintainable
solutions~\cite{dou2025wrong}---an expectation grounded in the view that
both correctness and quality reflect general coding competence. Our
results empirically disconfirm this assumption: a model atop a
correctness leaderboard may produce code harder to maintain, more deeply
nested, and more complex than a lower-ranked competitor. For SE
practitioners deploying LLMs in production, technical debt from
AI-generated code accumulates incrementally with real engineering
cost~\cite{perry2023insecure,dou2025wrong}. Evaluation frameworks must
treat correctness and quality as distinct, co-equal dimensions.
Figure~\ref{fig:ranking} illustrates this reversal directly, comparing
model rankings on correctness against rankings on quality score.

\begin{rqbox}
  \textbf{Answer to RQ3:} Functional correctness and composite code
  quality are empirically orthogonal (Pearson $r = 0.075$, $n = 340$)
  within this benchmark (85 C\# HumanEval tasks, four LLMs). The
  relationship is not statistically significant. This confirms that
  Pass@k rankings are insufficient as a sole evaluation criterion for
  LLMs in software engineering contexts; correctness and structural
  quality must be reported as independent dimensions. Whether this
  orthogonality generalises to other programming languages or task
  types remains an open question for future work.
\end{rqbox}

\subsection{GPT's Bimodal Failure Behavior}

\begin{figure}[ht]
  \centering
  \subcaptionbox{\footnotesize Model rankings by Correctness vs.\
    Quality Score, showing complete reversal ($r = 0.075$).\label{fig:ranking}}
    [0.48\linewidth]{\includegraphics[width=\linewidth]{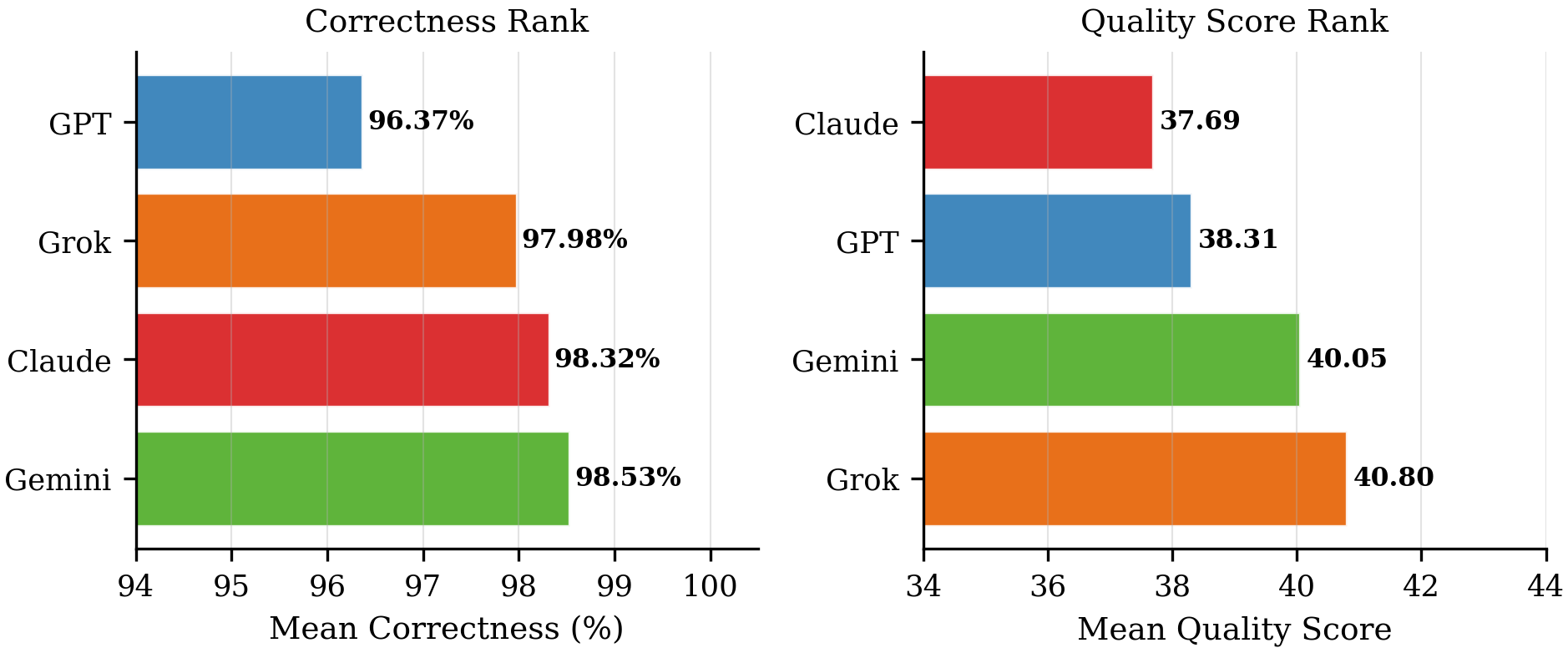}}
  \hfill
  \subcaptionbox{\footnotesize Quality Score on passed vs.\ failed tasks.
    GPT collapses 17.1 points on failures.\label{fig:bimodal}}
    [0.48\linewidth]{\includegraphics[width=\linewidth]{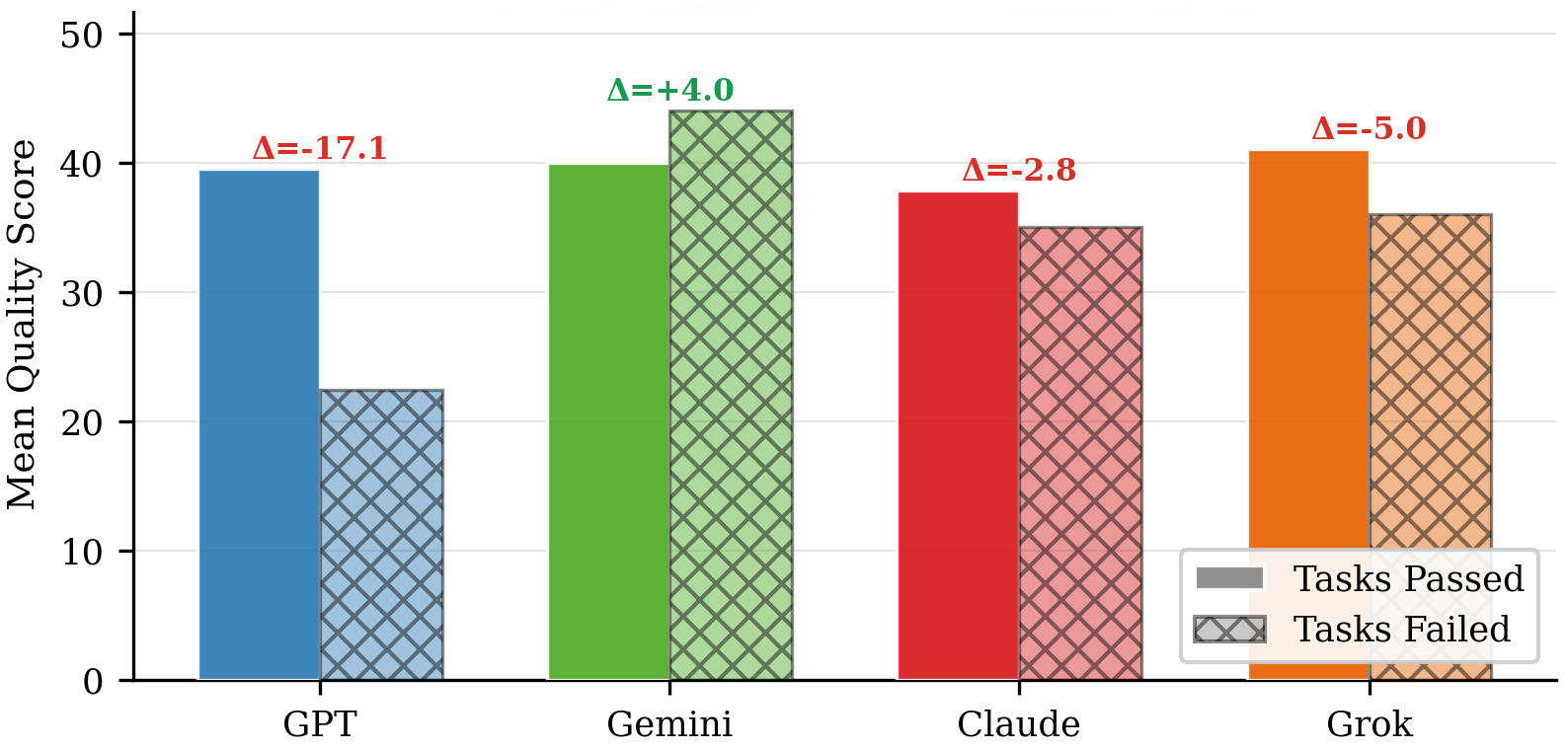}}
  \caption{\footnotesize Overall comparison of model rankings and bimodal
    quality scores.}
  \label{fig:both_charts}
\end{figure}

Figure~\ref{fig:bimodal} stratifies Quality Score by task outcome,
revealing GPT's pronounced quality collapse on failed tasks. GPT's
quality on failed tasks (22.44) is less than 57\% of its quality on
successful tasks (39.51), and just over half of Gemini's quality on
failures (44.00). When GPT fails a task, the generated code exhibits
broad structural deficiencies alongside the functional failure,
representing a systemic rather than marginal breakdown. The other three
models degrade more gracefully.

\begin{rqbox}
  \textbf{Answer to RQ1:} Gemini achieves the highest mean correctness
  (98.53\%, 2 failures) and is the most reliable model on this
  benchmark. Claude (98.32\%) and Grok (97.98\%) follow closely. GPT
  records the lowest mean (96.37\%) and the highest variance
  (SD = 15.37), indicating a bimodal distribution between fully correct
  and substantially failed solutions. All models converge on identical
  partial solutions for two tasks (\texttt{csharp\_57},
  \texttt{csharp\_64}), suggesting shared limitations across model
  families.
\end{rqbox}

\subsection{Task Difficulty as a Complexity Confound}

Cross-model complexity agreement on identical tasks ($r = 0.63$ to $0.69$
across model pairs) shows that inherent task difficulty, rather than
model-specific behavior, is the primary driver of cyclomatic complexity.
Problems requiring recursive logic or multi-branch control force elevated
complexity across all models. Benchmarks comparing raw complexity across
heterogeneous task sets without controlling for task difficulty will
produce biased results. Practitioners and benchmark designers should
therefore compare structural complexity metrics only within the same task,
or normalise against a task-level baseline.

\begin{rqbox}
  \textbf{Answer to RQ2:} The cross-model complexity correlation analysis
  ($r = 0.63$--$0.69$ across all model pairs on identical tasks) confirms
  that inherent task difficulty is the primary driver of structural
  complexity in LLM-generated code, with model-specific differences being
  secondary. Benchmarks that compare raw cyclomatic complexity or nesting
  depth across models without controlling for task-level difficulty will
  produce biased conclusions.
\end{rqbox}

\subsection{Holistic Performance Profiles}

\begin{figure}[h]
  \centering
  \includegraphics[width=0.35\linewidth]{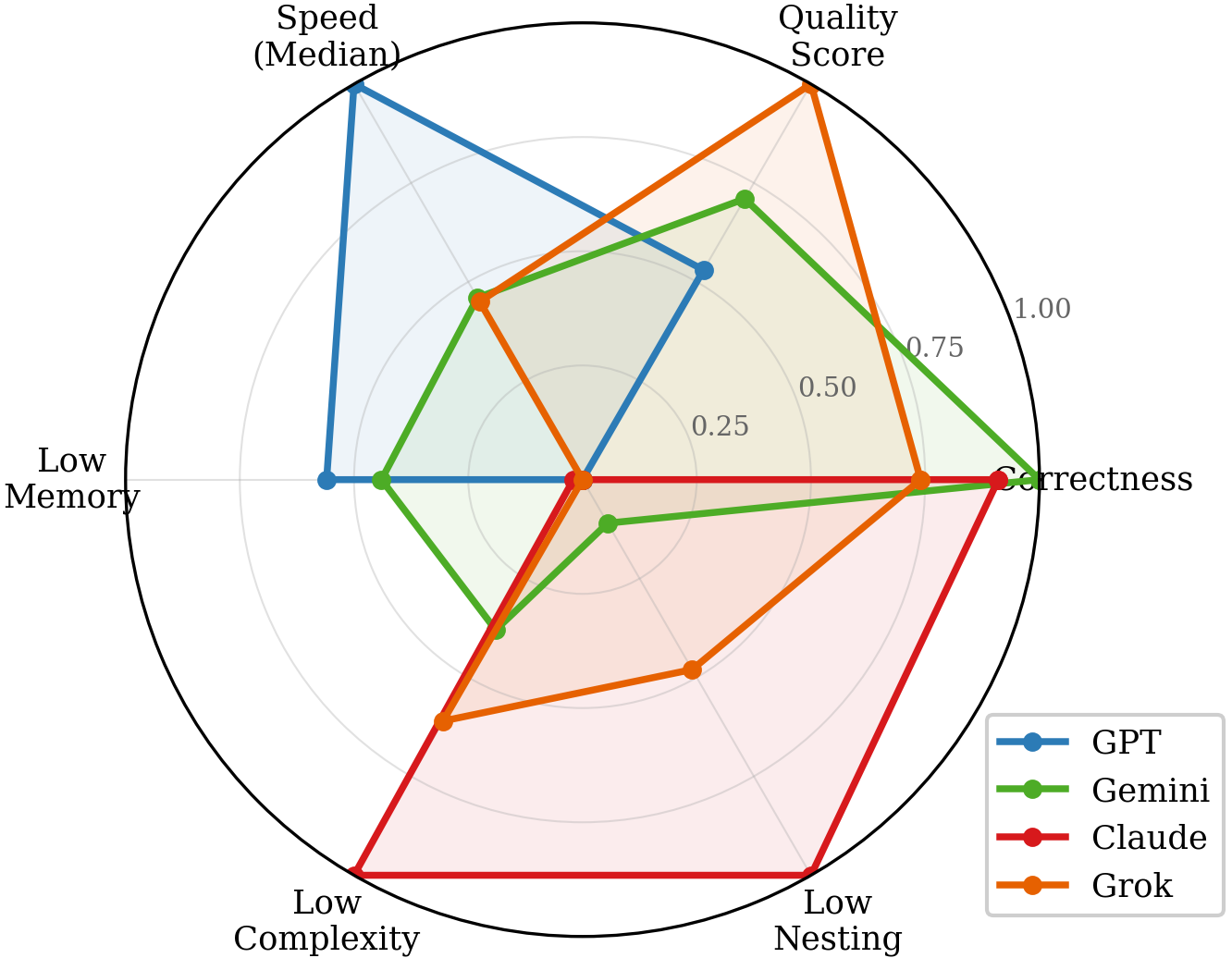}
  \caption{\footnotesize Normalized multi-dimensional radar across all
    six dimensions. Grok presents the most balanced profile.}
  \label{fig:radar}
\end{figure}

Figure~\ref{fig:radar} provides a normalized radar view across all six
evaluation dimensions simultaneously, making each model's trade-off
profile immediately visible. No model dominates across all dimensions.
Grok has the most balanced overall profile. Practitioners should select
models based on the quality dimension their context prioritizes, rather
than a single composite rank.

A noteworthy trade-off emerges from Claude's results: despite generating
the structurally simplest code (complexity 2.47, nesting 1.13), Claude
incurs the highest memory footprint (156.6\,MB median). This illustrates
that low structural complexity does not imply runtime efficiency---simpler
control flow can be achieved through data-intensive algorithmic strategies
that trade branch depth for memory cost. Correctness, structural
complexity, and runtime efficiency should therefore be treated as
complementary rather than redundant evaluation dimensions.

\begin{rqbox}
  \textbf{Answer to RQ4:} GPT is the fastest model (median 4,823\,ns);
  memory profile: Grok 6.0\,MB, GPT 11.7\,MB, Gemini 28.9\,MB, Claude
  156.6\,MB. Claude's outlier values in both dimensions are driven
  primarily by two anomalous tasks (\texttt{csharp\_12} and
  \texttt{csharp\_16}). All models remain within one order of magnitude
  on median execution time.
\end{rqbox}

\section{Threats to Validity}
\label{sec:threats}

We structure threats to validity following Wohlin et
al.~\cite{wohlin2012experimentation}.

\textbf{Internal validity.} LLM API non-determinism was mitigated by
setting temperature\,=\,0; all four providers document greedy decoding as
deterministic for a fixed model version. Multi-run variance analysis
under non-zero temperature is identified as future work. BenchmarkDotNet
outliers in Claude's runtime data were addressed by adopting the median
as the primary performance metric, with anomalous tasks
(\texttt{csharp\_12}, \texttt{csharp\_16}) explicitly identified.

\textbf{Construct validity.} The equal weighting of the five Composite
Quality Score sub-components may not reflect real-world priorities; all
decomposed metrics are reported individually to allow independent
interpretation. Since Norm\_C is one of the five components, the
$r = 0.075$ result is partly self-referential, specifically reflecting
the near-zero association between raw correctness and the four remaining
structural and performance components; a robustness check excluding
Norm\_C is identified as future work. The naming and professionalism
metrics capture implicit C\# convention adherence rather than
instruction-following, as style constraints were not included in the
solution prompt; investigating explicit style injection is also
identified as future work.

\textbf{External validity.} The 85-task HumanEval subset does not
reflect complex, multi-file enterprise .NET architectures, but provides a
necessary controlled baseline. The Python-to-C\# adaptation preserves
algorithmic intent; potential systematic difficulty bias is acknowledged
as an additional concern. The orthogonality finding (RQ3) should not be
assumed to generalise unconditionally to other languages or task types.
The framework is dataset-agnostic; applying it to enterprise repositories
is a key future direction.

\textbf{Conclusion validity.} The 85-task dataset may limit statistical
power; this was mitigated by performing the Pearson correlation analysis
over all 340 solution records and reporting standard deviations
throughout.

\section{Conclusion and Future Work}
\label{sec:conclusion}

This paper presents an automated, multi-dimensional evaluation framework
for C\# code generation. Our central contribution is demonstrating a
systematic gap between an LLM's ability to generate functionally correct
code and its capacity to produce structurally sound, maintainable, and
efficient code. By confirming that correctness and quality are virtually
orthogonal ($r = 0.075$, within this C\# HumanEval benchmark), our
findings directly challenge the sufficiency of Pass@k benchmarks for
selecting models in production software engineering.

Evaluation of four state-of-the-art LLMs revealed distinct performance
profiles: Gemini leads in functional reliability (98.53\% correctness);
Grok achieves the highest, most balanced composite quality score; Claude
generates the structurally simplest code but incurs the highest memory
footprint (156.6\,MB median), illustrating that structural simplicity and
runtime efficiency are independent properties; and GPT writes the
fastest-executing code but exhibits the highest structural complexity and
a severe, bimodal quality collapse on failed tasks. Furthermore, identical
partial solutions across all models suggest shared biases in their
underlying training distributions.

Future work will expand this framework to enterprise-representative C\#
patterns (e.g., asynchronous programming, LINQ pipelines), empirically
calibrate Quality Score weights via expert studies, investigate
prompt-injected style and naming constraints, conduct multi-run variance
analysis, and validate the RQ3 orthogonality finding with a
four-component quality score that excludes correctness.

\paragraph{Replication.}
The evaluation harness, adapted task specifications, and raw results
dataset are publicly available on
GitHub.\footnote{\url{https://github.com/mohammadmehdighalandarian/llm4se.git}}
All experiments are fully reproducible given the documented hardware
configuration and API temperature settings.

\section*{Declaration on Generative AI}
During the preparation of this work, the author(s) used Claude for writing assistance, grammar and spelling correction, and  \LaTeX{} formatting. The author(s) reviewed and edited the content as needed and take full responsibility for the publication’s content.

\newpage
{
\fontsize{9pt}{10.5pt}\selectfont
\setlength{\bibsep}{0pt}
\renewcommand{\baselinestretch}{0.95}\selectfont
\bibliography{references}

@inproceedings{svyatkovskiy2020intellicode,
  author    = {Svyatkovskiy, Alexey and Deng, Shao Kun and Fu, Shengyu and Sundaresan, Neel},
  title     = {{IntelliCode Compose: Code Generation Using Transformer}},
  booktitle = {Proceedings of the 28th ACM Joint European Software Engineering
               Conference and Symposium on the Foundations of Software Engineering
               (ESEC/FSE)},
  year      = {2020},
  pages     = {1433--1443},
  publisher = {ACM},
  doi       = {10.1145/3368089.3417058}
}

@inproceedings{ziegler2022productivity,
  author    = {Ziegler, Albert and Kalliamvakou, Eirini and Li, X. Alice and
               Rice, Andrew and Rifkin, Devon and Simister, Shawn and
               Sittampalam, Ganesh and Aftandilian, Edward},
  title     = {{Productivity Assessment of Neural Code Completion}},
  booktitle = {Proceedings of the 6th ACM SIGPLAN International Symposium on
               Machine Programming (MAPS)},
  year      = {2022},
  pages     = {21--29},
  publisher = {ACM}
}

@inproceedings{perry2023insecure,
  author    = {Perry, Neil and Srivastava, Megha and Kumar, Deepak and Boneh, Dan},
  title     = {{Do Users Write More Insecure Code with AI Assistants?}},
  booktitle = {Proceedings of the ACM SIGSAC Conference on Computer and
               Communications Security (CCS)},
  year      = {2023},
  pages     = {2785--2799},
  publisher = {ACM}
}

@inproceedings{pearce2022asleep,
  author    = {Pearce, Hammond and Ahmad, Baleegh and Tan, Benjamin and
               Dolan-Gavitt, Brendan and Karri, Ramesh},
  title     = {{Asleep at the Keyboard? Assessing the Security of GitHub
               Copilot's Code Contributions}},
  booktitle = {Proceedings of the IEEE Symposium on Security and Privacy (S\&P)},
  year      = {2022},
  pages     = {754--768},
  publisher = {IEEE}
}

@article{chen2021codex,
  author    = {Chen, Mark and Tworek, Jerry and Jun, Heewoo and Yuan, Qiming and
               Pinto, Henrique Ponde de Oliveira and Kaplan, Jared and
               Edwards, Harri and Burda, Yuri and Joseph, Nicholas and
               Brockman, Greg and others},
  title     = {{Evaluating Large Language Models Trained on Code}},
  journal   = {arXiv preprint arXiv:2107.03374},
  year      = {2021}
}

@article{austin2021mbpp,
  author    = {Austin, Jacob and Odena, Augustus and Nye, Maxwell and
               Bosma, Maarten and Michalewski, Henryk and Dohan, David and
               Jiang, Ellen and Cai, Carrie and Terry, Michael and
               Le, Quoc and others},
  title     = {{Program Synthesis with Large Language Models}},
  journal   = {arXiv preprint arXiv:2108.07732},
  year      = {2021}
}

@article{li2022alphacode,
  author    = {Li, Yujia and Choi, David and Chung, Junyoung and
               Kushman, Nate and Schrittwieser, Julian and Leblond, R{\'e}mi and
               Eccles, Tom and Keeling, James and Gimeno, Felix and
               Dal Lago, Agustin and others},
  title     = {{Competition-Level Code Generation with AlphaCode}},
  journal   = {Science},
  volume    = {378},
  number    = {6624},
  pages     = {1092--1097},
  year      = {2022},
  publisher = {AAAS}
}

@inproceedings{zheng2023codegeex,
  author    = {Zheng, Qinkai and Xia, Xiao and Zou, Xu and Dong, Yuxiao and
               Wang, Shan and Xue, Yufei and Wang, Zihan and Shen, Lei and
               Wang, Andi and Li, Yang and others},
  title     = {{CodeGeeX: A Pre-Trained Model for Code Generation with
               Multilingual Benchmarking on HumanEval-X}},
  booktitle = {Proceedings of the 29th ACM SIGKDD Conference on Knowledge
               Discovery and Data Mining},
  year      = {2023},
  pages     = {5673--5684},
  publisher = {ACM}
}

@inproceedings{paul2024benchmarks,
  author    = {Paul, Debalina Ghosh and Zhu, Hong and Bayley, Ian},
  title     = {{Benchmarks and Metrics for Evaluations of Code Generation:
               A Critical Review}},
  booktitle = {Proceedings of the IEEE International Conference on Artificial
               Intelligence Testing (AITest)},
  year      = {2024},
  pages     = {87--94},
  publisher = {IEEE},
  doi       = {10.1109/AITest62860.2024.00019}
}

@book{akinshin2019dotnet,
  author    = {Akinshin, Andrey},
  title     = {{Pro .NET Benchmarking: The Art of Performance Measurement}},
  publisher = {Apress},
  address   = {New York, NY, USA},
  year      = {2019}
}

@inproceedings{lian2024deveval,
  author    = {Lian, Jia and others},
  title     = {{DevEval: A Manually-Annotated Code Generation Benchmark
               Aligning with Real-World Code Repositories}},
  booktitle = {Proceedings of the 62nd Annual Meeting of the Association for
               Computational Linguistics (ACL)},
  year      = {2024},
  publisher = {ACL}
}

@inproceedings{zheng2025humanevo,
  author    = {Zheng, Dong and Wang, Yanlin and Shi, Ensheng and Zhang, Ruikai
               and Ma, Yuchi and Zhang, Hongyu and Zheng, Zibin},
  title     = {{HumanEvo: An Evolution-Aware Benchmark for More Realistic
               Evaluation of Repository-Level Code Generation}},
  booktitle = {Proceedings of the 47th IEEE/ACM International Conference on
               Software Engineering (ICSE)},
  year      = {2025},
  pages     = {1372--1384},
  publisher = {IEEE/ACM},
  doi       = {10.1109/ICSE55347.2025.00228}
}

@inproceedings{jiao2023evaluation,
  author    = {Jiao, Mengting and Yu, Tao and Li, Xuan and Qiu, Gengwei and
               Gu, Bin and Shen, Baosheng},
  title     = {{On the Evaluation of Neural Code Translation: Taxonomy and
               Benchmark}},
  booktitle = {Proceedings of the 38th IEEE/ACM International Conference on
               Automated Software Engineering (ASE)},
  year      = {2023},
  pages     = {1529--1541},
  publisher = {IEEE/ACM},
  doi       = {10.1109/ASE56229.2023.00114}
}

@article{dou2025wrong,
  author    = {Dou, Shihan and Jia, Haoxiang and Wu, Shenxi and Zheng, Huiyuan
               and Zhou, Weikang and Wu, Muling and Chai, Mingxu and Fan, Jinhao
               and Huang, Cheng and Tao, Yan and Liu, Yan and Zhou, Enyu and
               Zhang, Ming and Zhou, Yueming and Wu, Yuming and Zheng, Rui and
               Wen, Ming and Weng, Rongxiang and Wang, Jingang and Cai, Xipeng
               and Gui, Tao and Qiu, Xipeng and Zhang, Qi and Huang, Xuanjing},
  title     = {{What Is Wrong with Your Code Generated by Large Language Models?
               An Extensive Study}},
  journal   = {Science China Information Sciences},
  volume    = {69},
  year      = {2025},
  publisher = {Springer},
  doi       = {10.1007/s11432-025-4632-8}
}

@article{zhang2024hallucinations,
  author    = {Zhang, Zongyi and Wang, Chen and Wang, Yingyu and Shi, Ensheng
               and Ma, Yuchi and Zhong, Wanjun and Chen, Jichuan and Mao, Mingzhe
               and Zheng, Zibin},
  title     = {{LLM Hallucinations in Practical Code Generation: Phenomena,
               Mechanism, and Mitigation}},
  journal   = {Proceedings of the ACM on Software Engineering},
  volume    = {2},
  pages     = {481--503},
  year      = {2024},
  doi       = {10.1145/3728894}
}

@article{moudgalya2023tasty,
  author    = {Moudgalya, Kaushik and Ramakrishnan, Ankit and
               Chemudupati, Vamsikrishna and Lu, Xing Han},
  title     = {{TASTY: A Transformer based Approach to Space and Time
               Complexity}},
  journal   = {arXiv preprint arXiv:2305.05379},
  year      = {2023},
  doi       = {10.48550/arXiv.2305.05379}
}

@article{mccabe1976complexity,
  author    = {McCabe, Thomas J.},
  title     = {{A Complexity Measure}},
  journal   = {IEEE Transactions on Software Engineering},
  volume    = {SE-2},
  number    = {4},
  pages     = {308--320},
  year      = {1976},
  publisher = {IEEE}
}

@article{mccabe1989design,
  author    = {McCabe, Thomas J. and Butler, Charles W.},
  title     = {{Design Complexity Measurement and Testing}},
  journal   = {Communications of the ACM},
  volume    = {32},
  number    = {12},
  pages     = {1415--1425},
  year      = {1989},
  publisher = {ACM},
  doi       = {10.1145/76380.76382}
}

@article{huang2023empirical,
  author    = {Huang, Qing and Sun, Jiahao and Xu, Xin and Hu, Qiao and
               Liu, Jia and Luo, Xiapu},
  title     = {{An Empirical Study on Challenges and Issues of Code Generation
               with Large Language Models}},
  journal   = {arXiv preprint arXiv:2310.06218},
  year      = {2023}
}

@inproceedings{kalibera2013rigorous,
  author    = {Kalibera, Tomas and Jones, Richard},
  title     = {{Rigorous Benchmarking in Reasonable Time}},
  booktitle = {Proceedings of the ACM SIGPLAN International Symposium on
               Memory Management (ISMM)},
  year      = {2013},
  pages     = {63--74},
  publisher = {ACM}
}

@article{deepseek2024v3,
  author    = {{DeepSeek-AI}},
  title     = {{DeepSeek-V3 Technical Report}},
  journal   = {arXiv preprint arXiv:2412.19437},
  year      = {2024},
  doi       = {10.48550/arXiv.2412.19437}
}

@article{liang2022helm,
  author    = {Liang, Percy and Bommasani, Rishi and Lee, Tony and Tsipras,
               Dimitris and Soylu, Dilara and Yasunaga, Michihiro and Zhang,
               Yian and Narayanan, Deepak and Wu, Yuhuai and Kumar, Ananya and
               others},
  title     = {{Holistic Evaluation of Language Models}},
  journal   = {arXiv preprint arXiv:2211.09110},
  year      = {2022}
}

@article{chang2024survey,
  author    = {Chang, Yupeng and Wang, Xu and Wang, Jindong and Wu, Yuan and
               Yang, Linyi and Zhu, Kaijie and Chen, Hao and Yi, Xiaoyuan and
               Wang, Cunxiang and Wang, Yidong and others},
  title     = {{A Survey on Evaluation of Large Language Models}},
  journal   = {ACM Transactions on Intelligent Systems and Technology},
  volume    = {15},
  number    = {3},
  pages     = {1--45},
  year      = {2024},
  publisher = {ACM}
}

@inproceedings{georges2007statistically,
  author    = {Georges, Andy and Buytaert, Dries and Eeckhout, Lieven},
  title     = {{Statistically Rigorous Java Performance Evaluation}},
  booktitle = {Proceedings of the 22nd ACM SIGPLAN Conference on
               Object-Oriented Programming, Systems, Languages and
               Applications (OOPSLA)},
  year      = {2007},
  pages     = {57--76},
  publisher = {ACM}
}

@article{openai2023gpt4,
  author    = {{OpenAI}},
  title     = {{GPT-4 Technical Report}},
  journal   = {arXiv preprint arXiv:2303.08774},
  year      = {2023},
  doi       = {10.48550/arXiv.2303.08774}
}

@article{google2024gemini,
  author    = {{Gemini Team}},
  title     = {{Gemini 1.5: Unlocking Multimodal Understanding Across Millions
               of Tokens of Context}},
  journal   = {arXiv preprint arXiv:2403.05530},
  year      = {2024},
  doi       = {10.48550/arXiv.2403.05530}
}

@techreport{anthropic2024claude,
  author      = {{Anthropic}},
  title       = {{Claude 3.5 Sonnet Model Card}},
  institution = {Anthropic},
  year        = {2024},
  url         = {https://www.anthropic.com/claude/sonnet}
}

@techreport{xai2025grok,
  author      = {{xAI}},
  title       = {{Grok-3 Technical Report}},
  institution = {xAI},
  year        = {2025},
  url         = {https://x.ai/news/grok-3}
}

@article{kitchenham2002preliminary,
  author    = {Kitchenham, Barbara A. and Pfleeger, Shari Lawrence and
               Pickard, Lesley M. and Jones, Peter W. and Hoaglin, David C.
               and El Emam, Khaled and Rosenberg, Jarrett},
  title     = {{Preliminary Guidelines for Empirical Research in Software
               Engineering}},
  journal   = {IEEE Transactions on Software Engineering},
  volume    = {28},
  number    = {8},
  pages     = {721--734},
  year      = {2002},
  publisher = {IEEE}
}

@book{wohlin2012experimentation,
  author    = {Wohlin, Claes and Runeson, Per and H{\"o}st, Martin and
               Ohlsson, Magnus C. and Regnell, Bj{\"o}rn and
               Wessl{\'e}n, Anders},
  title     = {{Experimentation in Software Engineering}},
  publisher = {Springer},
  address   = {Berlin, Heidelberg},
  year      = {2012}
}
}

\end{document}